\documentclass[aps,prapplied,superscriptaddress,twocolumn,floatfix,nofootinbib,nobibnotes,10pt]{revtex4-2}
\usepackage{graphicx}
\usepackage{amsmath,amssymb,amsfonts}
\usepackage{bbold}
\usepackage{algorithmic}
\usepackage{gensymb}
\usepackage{fancyhdr}
\usepackage[colorlinks,citecolor=blue,linkcolor=blue,urlcolor=blue]{hyperref}

\newcommand{\ket}[1]{|#1\rangle}
\newcommand{\bra}[1]{\langle#1|}
\newcommand{\braket}[2]{\langle#1|#2\rangle}
\newcommand{\ensavg}[1]{\langle#1\rangle}

\makeatletter
\def\maketag@@@#1{\hbox{\m@th\normalfont\normalsize#1}}  
\makeatother

\begin{document}
\hypersetup{pageanchor=false}
\title{Satellite-to-Earth Quantum Key Distribution via Orbital Angular Momentum}

\author{Ziqing Wang}\email{ziqing.wang1@unsw.edu.au}
\affiliation{School of Electrical Engineering and Telecommunications, University of New South Wales, Sydney, NSW 2052, Australia}

\author{Robert Malaney}\email{r.malaney@unsw.edu.au}
\affiliation{School of Electrical Engineering and Telecommunications, University of New South Wales, Sydney, NSW 2052, Australia}

\author{Benjamin Burnett}
\affiliation{Northrop Grumman Corporation, San Diego, California, USA}

\date{\today}

\begin{abstract}
In this work, we explore the feasibility of performing satellite-to-Earth quantum key distribution (QKD) using the orbital angular momentum (OAM) of light. Due to the fragility of OAM states the conventional wisdom is that turbulence would render OAM-QKD non-viable in a satellite-to-Earth channel. However, based on detailed phase screen simulations of the anticipated atmospheric turbulence we find that OAM-QKD is viable in some system configurations, especially if quantum channel information is utilized in the processing of post-selected states. More specifically, using classically entangled light as a probe of the quantum channel, and reasonably-sized  transmitter-receiver apertures,  we find that non-zero QKD rates are achievable on sea-level ground stations. Without using classical light probes, OAM-QKD is relegated to high-altitude ground stations with large receiver apertures. Our work represents the first quantitative assessment of the performance of OAM-QKD from satellites, showing under what circumstances the much-touted higher dimensionality of OAM can be utilized in the context of secure communications.
\end{abstract}

\maketitle
\thispagestyle{fancy}
\renewcommand{\headrulewidth}{0pt}
\fancyfoot[C]{}
\fancyhead[R]{\ifthenelse{\value{page}=1}{}{\thepage}}
\fancyhead[L]{}

\section{Introduction}
As one of the most important applications in quantum communications, Quantum Key Distribution (QKD) has been proven to provide unconditional security~\cite{BB84}. Recently, real-world implementations of satellite-based QKD (e.g.~\cite{SatQKD,ChinaSatEBQKD}) have pointed the way towards global-scale and highly-secure quantum communication networks~\cite{sat-based_QNet}. 
The originally proposed QKD protocols (e.g.~\cite{BB84,E91,SSP}) mainly utilize 2-dimensional encoding. However, other QKD protocols have been generalized to the case of high-dimensional encoding (e.g.~\cite{QKD_Multilevel}), and their unconditional security has been proved (e.g.~\cite{QKD_Multilevel_Security_JPA,QKD_Multilevel_Security_PRL,Qudit_SecurityProof2010,Qudit_SecurityProof2012}).
Quantum information can be encoded in any degree of freedom (DoF) of the photon, but most of the mainstream implementations of QKD (e.g.~\cite{ChinaSatEBQKD,SatQKD,sat-based_QNet}) rely on polarization encoding - a typical 2-dimensional encoding scheme that limits the capacity of QKD systems due to an intrinsically bounded Hilbert space.

The Orbital Angular Momentum (OAM) of light has been considered as a promising DoF for quantum communications~\cite{OAM_3D_Entanglement}. Unlike the polarization of light, the OAM of light can take arbitrary integer values~\cite{Allen92}. The corresponding OAM eigenstates form an orthonormal basis that allows for quantum coding within a theoretically infinite-dimensional Hilbert space, opening up new possibilities for high-capacity quantum communications. As a key resource for quantum communications, entanglement can be encoded in OAM via the spontaneous parametric down-conversion (SPDC) process~\cite{OAM_entanglement,OAM_entanglement12}. The distribution of OAM-encoded entanglement through the turbulent atmosphere has been intensively investigated in \emph{terrestrial} free-space optical (FSO) channels (e.g.~\cite{OAM_Entanglement_SPS_Theory,OAM_Entanglement_SPS,OAM_Entanglement_IPE_Theory,OAM_Entanglement_MPS,EntanglementProtectionAO2018,EntanglementProtectionAO2019}) with some demonstrating distribution over 3~km~\cite{OAM_entanglement_3km}. A recent experiment suggests that OAM entanglement distribution could be feasible over an FSO channel of more than 100 km~\cite{OAM_143km}.

Besides the generation and distribution of OAM-encoded entanglement, other recent efforts have paved the way for the practical implementation of OAM-QKD. Any OAM superposition state can be efficiently encoded in single photons thanks to the versatility of a spatial light modulator (SLM) (see e.g.~\cite{OAM_Gen_Superposition_2013,OAM_2MUB_Generation_2013}). The sorting of OAM-photons has also been made possible (e.g.~\cite{OAMSorter2013,GeneralizedOAMSorter2017,OAM_FullSort_2020}), enabling the capability of performing multi-outcome measurements. Implementations of OAM-QKD have been demonstrated in laboratory conditions with 2-dimensional (e.g.~\cite{OAM_QKD_Many_2018}) and higher-dimensional (e.g.~\cite{OAM_QKD_Qutrit_2006,OAM_QKD_Qudit_2013,OAM_QKD_Qudit_2015,OAM_QKD_Many_2018}) encoding. Efforts have also been made to investigate the practical feasibility of performing OAM-QKD in turbulent \emph{terrestrial} FSO  channels~\cite{OAM_SSP,OAM_QKD_Qubit_TurbBoth2016}. Outside the laboratory, OAM-QKD has been demonstrated over turbulent FSO channels of 210 m~\cite{OAM_QKD_Qubit_RI_2014}, and 300 m~\cite{OAM_QKD_Qubit_Hybrid2017}. Considering other types of medium, OAM-QKD has also been demonstrated over a 3 m underwater link~\cite{OAM_QKD_Underwater2018} and a 1.2 km optical fiber~\cite{OAM_QKD_Fiber_2019}. However, most existing research on OAM-QKD has not considered the context of a \emph{satellite-based deployment}. As such, the feasibility of long-range OAM-QKD via satellite is still not clear.

Previously we have studied the OAM detection performance in satellite-to-Earth communications~\cite{ZiqingOAMDetection}, and the feasibility of OAM-based entanglement distribution via satellite~\cite{ZiqingOAMEntDist}. In this work, we  explore the feasibility of satellite-to-Earth OAM-QKD.
Our main finding is that, contrary to conventional wisdom, such QKD is indeed feasible.
More specifically, we find that utilizing quantum channel information
enables satellite-to-Earth OAM-QKD over a wide range of dimensions  under all anticipated circumstances, including the circumstance where a sea-level ground station with a reasonably-sized receiver aperture is used. If channel information is not used, then feasible satellite-to-Earth OAM-QKD is confined to large telescopes situated at high-altitude observatories.

The remainder of this paper is as follows.
In Section~\ref{Sec:Background} we introduce the necessary background knowledge on OAM eigenstates, atmospheric propagation of light, and the generalized OAM-QKD protocol.
In Section~\ref{Sec:SystemModel} we detail the system model for satellite-to-Earth OAM-QKD.
In Section~\ref{Sec:Simulation} we present our key results on satellite-to-Earth OAM-QKD.
In Section~\ref{Sec:Distillation} we explore the use of quantum channel information to improve the practical feasibility of satellite-to-Earth OAM-QKD.
Finally, concluding remarks are provided in Section~\ref{Sec:Conclusion}.

\section{Background}\label{Sec:Background}
\subsection{OAM eigenstates}
OAM-QKD protocols utilize OAM eigenstates and their superpositions for quantum encoding. In cylindrical coordinates, the general form of an OAM eigenstate is given by
\begin{equation}\label{Eq.OAM_Eigenstate}
\varphi _ { p , l } ( r , \theta,z) = R _ {p,l} ( r,z ) \frac{ \exp ( i l \theta ) }{\sqrt{2\pi}},
\end{equation}
where $r$ and $\theta$ are the radial and azimuthal coordinates, respectively, $z$ is the longitudinal distance, $l$ is the OAM quantum number, $p$ is the radial node number, and $ R_{p,l} (r,z)$ is the radial profile. OAM eigenstates with different $l$ values are mutually orthogonal. In this paper, we choose $ R_{p,l} (r,z)$ to be Laguerre-Gauss functions, making OAM eigenstates correspond to the Laguerre-Gaussian (LG) mode set~\cite{OAM_Rev11}. $R_{p,l} (r,z)$ is expressed in Eq.~(\ref{LG}) as
\begin{equation}\label{LG}
\begin{aligned} R_{p,l}(r,z)= \ &2 \sqrt{\frac{p !}{(p+|l|) !}} \frac { 1 } { w ( z ) } \left[ \frac { r \sqrt { 2 } } { w ( z ) } \right] ^ { | l | } \exp \left[ \frac { - r ^ { 2 } } { w ^ { 2 } ( z ) } \right] \\
& L _ { p } ^ { |l|} \left( \frac { 2 r ^ { 2 } } { w ^ { 2 } ( z ) } \right) \exp \left[ \frac { i k r ^ { 2 } z } { 2 \left( z ^ { 2 } + z _ { R } ^ { 2 } \right) } \right] \\
& \exp \left[ - i ( 2 p + | l | + 1 ) \tan ^ { - 1 } \left( \frac { z } { z _ { R } } \right) \right],
\end{aligned}
\end{equation}
where $w(z)=w_{0} \sqrt{1+(z / z_{R})^{2}}$, $w_0$ is the beam-waist radius, $z_{\mathrm{R}}=\pi w_{0}^{2}/\lambda$ is the Rayleigh range, $\lambda$ is the optical wavelength,  $k=2\pi/\lambda$ is the optical wavenumber, and $L_{p}^{|l|}(x)$ is the generalized Laguerre polynomial. We denote the single-photon OAM eigenstate of the $\text{LG}_{pl}$ mode as $|p l\rangle$, and this notation is further simplified to $|l\rangle$ as we only consider the $p=0$ subspace. We denote the set $\{\ket{l}, \,-\infty<l<\infty\}$ as the OAM basis and use it as the standard basis throughout this paper. Denoting the dimension as $d$, the standard basis of $d$-dimensional OAM-QKD contains $d$ mutually orthogonal OAM eigenstates and thus spans a $d$-dimensional Hilbert space.
Throughout this work we will denote such a $d$-dimensional Hilbert space as the encoding subspace $\mathcal{H}_d$.

In this work, we consider a maximum OAM number of 4 to construct the encoding subspace $\mathcal{H}_d$. Specifically, we use the same approach adopted in~\cite{EntanglementProtectionAO2019} to construct the encoding subspace  $\mathcal{H}_d$. For $d=2$, we consider a 2-dimensional encoding subspace spanned by a pair of OAM eigenstates with opposite OAM numbers (i.e. $\mathcal{H}_2 = \{-l_0,l_0\}$ with $l_0\le4$). For $d=3$, we consider a 3-dimensional encoding subspace spanned by a pair of OAM eigenstates with opposite OAM numbers and the OAM eigenstate with zero OAM number (i.e. $\mathcal{H}_3 = \{-l_0,0,l_0\}$ with $l_0\le4$). For $d>3$, more OAM numbers are involved. For example, for $d=4$ the 4-dimensional encoding subspace is spanned by two pairs of OAM eigenstates with opposite OAM numbers (i.e. $\mathcal{H}_4=\{-l_2,-l_1,l_1,l_2\}$ with $l_1<l_2\le4$).


\subsection{Mutually unbiased bases}\label{SubSec:MUBs}
Denoted by $\mathcal{M_{\beta}}=\{\ket{\xi_{(\beta,s)}},\beta=1,\dots,d+1,\,s=0,\dots,d-1\}$, mutually unbiased bases (MUBs) are orthonormal bases defined on a $d$-dimensional Hilbert space such that
\begin{equation}\label{Eq:MUB}
\left|\braket{\xi_{(\beta,s)}}{\xi_{(\beta^{\prime},s^{\prime})}}\right|^2=\left\{\begin{array}{ll}{\delta_{s,s'}} & {\text { if } \beta=\beta'} \\ {\frac{1}{{d}}} & {\text { if } \beta \neq \beta^{\prime}}\end{array},\right.
\end{equation}
where $\delta$ denotes the Kronecker delta function.
MUBs play an important role in QKD since any system prepared in a state in one MUB gives outcomes with equal probability $1/d$ if measured in any other MUB. Therefore, if the eavesdropper measures the quantum signal in a wrong basis, she will acquire no information (in fact, she will introduce a disturbance).



It has been proven that, for a prime-power dimension $d$ there exists a complete set of $d+1$ MUBs~\cite{MUB_New_Proof}. In this work we consider a variety of dimensions ranging from $d=2$ to $d=9$. When $d$ is a prime number (i.e. $2$, $3$, $5$, $7$ in this work), a complete set of $d+1$ MUBs is found as eigenstates of different Weyl operators in the set $\{Z,XZ^{n}| \,n =0,1, \dots, d-1\}$. The $Z$ operator is defined as
\begin{equation}\label{Eq:Z_Operator}
Z=\sum_{j=0}^{d-1} \vartheta^{j}\ket{j}\bra{j},
\end{equation}
where $\ket{j}$ denote the standard basis elements, and $\vartheta=\exp{(i2\pi/d)}$. The $X$ operator is defined as
\begin{equation}\label{Eq:X_Operator}
X=\sum_{j=0}^{d-1}|(j+1) \bmod d\rangle\langle j|.
\end{equation}
When $d$ is a prime-power number but not a prime number (i.e. $4$, $8$, $9$ in this work), the construction of a complete set of $d+1$ MUBs becomes a harder task. In this work we adopt the sets of MUBs given in~\cite{MUB_PrimePower,MUB_PrimePower_GraphState} for these dimensions. The only non-prime-power dimension considered in this work is $d=6$. Since the maximum number of MUBs is not known for an arbitrary dimension, for $d=6$ we  use only the 2 MUBs generated from the set $\{Z,\, X\}$ (note that this has a negligible impact on the findings of this work).
In OAM-QKD, the standard basis is the OAM basis, thus any $\ket{\xi_{(\beta,s)}}$ is a superposition of OAM eigenstates that span the encoding subspace $\mathcal{H}_d$.

\subsection{Optical propagation through turbulent atmosphere}
\label{SubSec:TurbProp}
The turbulent atmosphere is a random medium with random inhomogeneities (turbulent eddies) of different size scales that are upper-bounded and lower-bounded by an outer scale $L_{\text{outer}}$ and an inner scale $l_{\text{inner}}$, respectively. These turbulent eddies give rise to small random refractive index fluctuations, causing continuous phase modulations on the optical beam. This leads to random refraction and diffraction effects, imposing distortions on the optical beam as it propagates through the atmospheric channel.

Under the paraxial approximation, the propagation of a monochromatic optical beam $\psi$ through the turbulent atmosphere is governed by the stochastic parabolic equation~\cite{book}
\begin{equation}\label{sHe}
\nabla_{\text{T}}^{2} \psi(\mathbf{R})+i 2 k \partial_{z} \psi(\mathbf{R})+2 \delta n(\mathbf{R}) k^{2} \psi(\mathbf{R})=0,
\end{equation}
where $\mathbf{R}=[x,y,z]^{T}$ denotes the three-dimensional position vector (in Eq.~(\ref{sHe}) we use Cartesian coordinates for simplicity), $\nabla_{\text{T}}^{2}=\partial^{2} / \partial x^{2}+\partial^{2} / \partial y^{2}$ is the transverse Laplacian operator, and $ \delta n(\mathbf{R})\!=\!n(\mathbf{R})\!-\!\langle n(\mathbf{R})\rangle$ represents the small refractive index fluctuation, with $n(\mathbf{R})$ being the refractive index at $\mathbf{R}$. Note that the turbulent atmosphere satisfies $\langle n(\mathbf{R}) \rangle\!=\!1$ and $\delta n(\mathbf{R})\!\ll\!1$~\cite{book}. In this work we numerically solve Eq.~(\ref{sHe}) using the \textit{split-step} method~\cite{Martin88,SSM1,SimAP}, which has been widely used to study atmospheric optical propagation under a variety of conditions.
This method models the atmospheric channel using multiple slabs with a phase screen located in the midway of each slab. Two free-space (vacuum) propagations  with one random phase modulation in between are repeatedly performed for each slab until the beam reaches the receiver plane~\cite{SSM1}. The split-step method has also been used to study the entanglement evolution of OAM-photon pairs in horizontal atmospheric channels, providing quantitative agreement with analytical results~\cite{OAM_Entanglement_MPS,EntanglementProtectionAO2018}.


\subsection{Generalized OAM-QKD protocol}\label{SubSec:Gen_QKD_Protocol}
QKD protocols can be described and implemented in both the prepare-and-measurement (P\&M) paradigm and the entanglement-based (EB) paradigm. Although most implementations of QKD are based on the P\&M scheme, all P\&M QKD protocols have their EB equivalences (note that EB QKD has also been demonstrated over the satellite-to-Earth channel, see e.g.~\cite{ChinaSatEBQKD}). Furthermore, the EB paradigm is usually adopted to simplify the security analysis. Throughout this work we adopt the EB paradigm for OAM-QKD. Here we briefly recall the procedures of a $d$-dimensional OAM-QKD protocol utilizing $N_B$ $(N_B \ge 2)$ MUBs.

\begin{enumerate}
	\item Alice first generates entangled photon pairs. For every pair of the entangled photons Alice keeps one photon at her side and sends the other photon to Bob through a quantum channel.
	
	\item For every photon pair, Alice and Bob randomly (and independently) choose one of the $N_B$ MUBs and perform a $d$-outcome measurement on their corresponding photon, giving each of them  a $d$-ary symbol.

	\item Alice and Bob start the sifting process where they reveal the MUBs that they used for their photon measurements. Specifically, they generate a sifted key by only keeping the symbols from the photon pairs jointly measured in the same MUB.
	
	\item {In the parameter estimation process, Alice and Bob compare a small subset of their sifted data to estimate the average error rate $Q$.}
	
	\item {With the knowledge on $Q$, the two parties then carry out subsequent processes, including reconciliation (which mainly includes error correction) and privacy amplification, to produce a final secret key that Eve has no knowledge on.}
\end{enumerate}

\section{System Model}\label{Sec:SystemModel}
\subsection{System settings}
\begin{figure}[hbt!]
	\hspace{8pt}
	\includegraphics[scale=0.55]{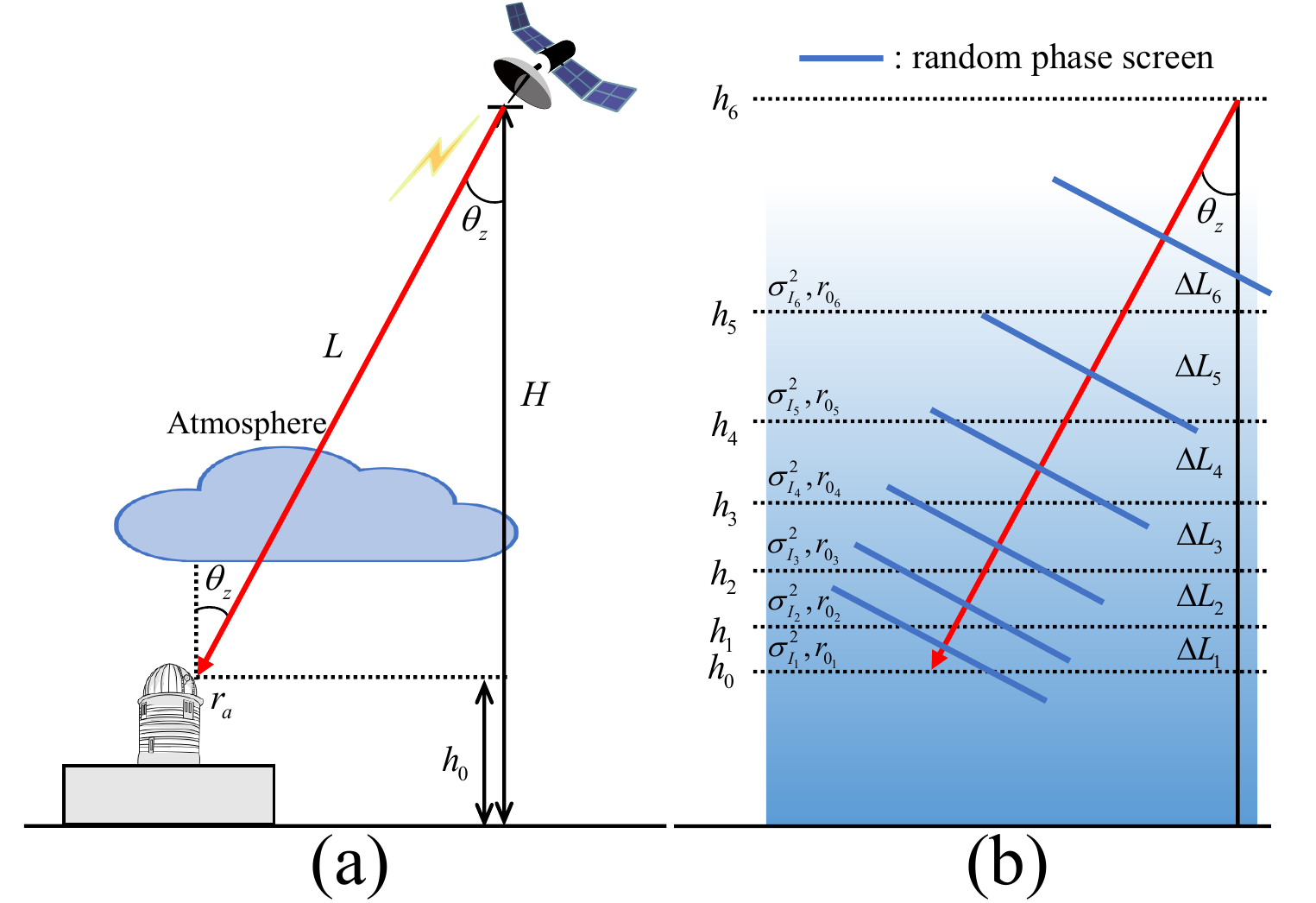}
	\caption{(a) System model for satellite-to-Earth OAM-QKD. (b) The modeling of a satellite-to-Earth atmospheric channel.}
	\label{fig:modelS}
\end{figure}

Throughout this work we denote the satellite and the ground station as Alice and Bob, respectively. In this section, we describe the system settings for  satellite-to-Earth OAM-QKD (as illustrated in Fig.~\ref{fig:modelS}(a)). The ground-station altitude is denoted as $h_0$, the satellite zenith angle at the ground station is denoted as $\theta_z$, and the satellite altitude at $\theta_z\!=\!0$ is denoted as $H$. The channel distance $L$ is given by $L\!=\!(H\!-\!h_0)\!/\!\cos\theta_z$. We denote the aperture radius at the ground station receiver as $r_{a}$. To perform OAM-QKD, Alice is equipped with an on-board SPDC source that generates entangled OAM-photon pairs. Both Alice and Bob are equipped with versatile OAM mode sorters that can randomly switch between all available MUBs and perform the corresponding $d$-outcome measurements.

\begin{figure}[hbt!]
	\centering
	\includegraphics[scale=0.4]{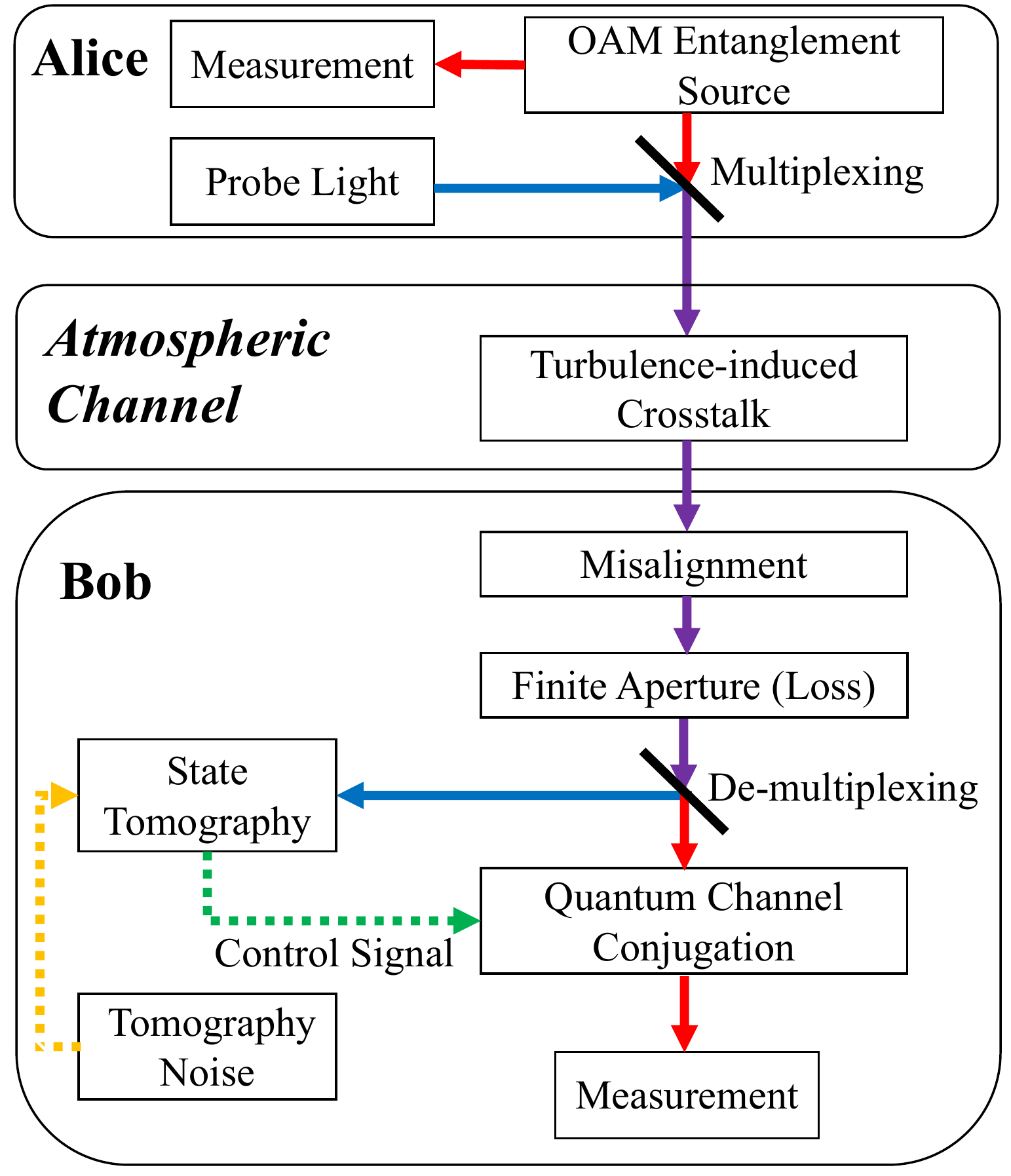}
	\caption{Schematic diagram for satellite-to-Earth OAM-QKD.}
	\label{fig:Schematic}
\end{figure}

The schematic diagram in Fig.~\ref{fig:Schematic} illustrates our deployment strategy for satellite-to-Earth OAM-QKD, in addition to all effects we consider. These include turbulence-induced crosstalk, loss (due to a finite-sized aperture), misalignment (due to imperfect beam tracking), and tomography noise (which leads to imperfect channel conjugation  when classically entangled light is used as a probe to characterize the quantum channel).

Unless otherwise specified, the following assumptions are adopted throughout this work:
\begin{enumerate}
	\item {We assume that Alice and Bob are perfectly time-synchronized, and they will discard any event where the photon sent by Alice does not click any of Bob's detectors.}
	\item{We assume that the OAM mode sorters used for measurement have a separation efficiency of unity and introduce no additional loss.}
	\item{For any specific dimension $d$, we only consider OAM-QKD protocols utilizing all $(d+1)$ MUBs. We restrict ourselves to the infinite key limit, therefore the sifting efficiency is set to 1. We also assume a reconciliation efficiency of 1.}
	\item{In the security analysis we assume that Eve controls the quantum channel and performs a collective attack.}
\end{enumerate}

\subsection{Satellite-to-Earth atmospheric channel}
\label{SubSec:DownlinkChannel}
\subsubsection{Turbulence characterization}
The strength of the optical turbulence within a satellite-based atmospheric channel can be described by the structure parameter $C_n^2(h)$ as a function of altitude $h$. $C_n^2(h)$ can be described by the widely used Hufnagel-Valley (HV) model~\cite{book}
\begin{equation}\label{Eq.HV}
\begin{aligned}
C_{n}^{2}(h)&\!=\!0.00594(v_{\text{rms}}/27)^{2}(h \times 10^{-5})^{10} \exp{(-h/1000)} \\
& +2.7\!\times\!10^{-16} \exp{(-h/1500)}\!+\!A \exp{(-h/100)},
\end{aligned}
\end{equation}
where $A$ is the ground-level (i.e. sea-level, $h=0$) turbulence strength in $\text{m}^{-2 / 3}$. In the above equation, $v_{\text{rms}}$ is the root-mean-square wind speed in m/s which is given by
\begin{equation}\label{Eq.Wind_Profile}
v_{\text{rms}}=\left[\frac{1}{15 \times 10^{3}} \int_{5 \times 10^{3}}^{20 \times 10^{3}} V^{2}(h) d h\right]^{1 / 2},
\end{equation}
where $V(h)$ is the altitude-dependent wind speed profile. In this paper we adopt the Bufton wind speed profile~\cite{book}
\begin{equation}\label{Bufton}
V(h)=V_g+30 \exp \left[-\left(\frac{h-9400}{4800}\right)^{2}\right],
\end{equation}
where $V_g$ is the ground-level wind speed.

The effect of the atmospheric turbulence on a propagating beam is quantified by two parameters, namely the scintillation index $\sigma_{I}^2$ and the Fried parameter $r_0$. The scintillation index is the normalized variance of the intensity. For satellite-to-Earth channels under weak-to-strong turbulence, this parameter is given by~\cite{book}
\begin{equation}\label{Eq.ScintIdxTot}
\sigma_{I}^{2}=\exp\!\left[\frac{0.49 \sigma_{R}^{2}}{\left(1+1.11 \sigma_{R}^{12 / 5}\right)^{7 / 6}}\!+\!\frac{0.51 \sigma_{R}^{2}}{\left(1+0.69 \sigma_{R}^{12 / 5}\right)^{5 / 6}}\right]\!-\!1,
\end{equation}
with $\sigma_{R}^2$ being the Rytov variance,
\begin{equation}\label{RytovVar}
\sigma_{R}^{2}=2.25 k^{7 / 6} \sec ^{11 / 6}(\theta_z) \int_{h_{0}}^{H} C_{n}^{2}(h)\left(h-h_{0}\right)^{5 / 6} d h.
\end{equation}
The Fried parameter quantifies the coherence length of the turbulence-induced phase errors in the transverse plane. For satellite-to-Earth channels, this parameter is given by~\cite{book}
\begin{equation}\label{Eq.FriedParamTot}
r_0=\left[0.423 k^{2} \sec \theta_z \int_{h_0}^{H} C_{n}^{2}(h) d h\right]^{-3 / 5}.
\end{equation}

\subsubsection{Channel modeling}
To perform the split-step method we divide the satellite-to-Earth atmospheric channel into $N_S$ slabs bounded by specific altitudes $h_j$ with $j$ ranging from 1 to $N_S$ (note that $h_0$ is the ground-station altitude, and a larger $j$ indicates a higher altitude). For the $j^{\text{th}}(j\!\ge\!1)$ slab bounded by $h_j$ and $h_{j-1}$, its thickness can be estimated as $\Delta L_j\!=\!(h_j\!-\!h_{j-1})\!/\!\cos(\theta_z)$ (note that $\sum_{j}\!\Delta L_{j}\!=\!L$). In order to characterize the turbulence within each slab, both $\sigma_{I}^{2}$ and $r_0$ are evaluated locally (for the turbulent volume of their corresponding slab). We denote the scintillation index and the Fried parameter for the $j^{\text{th}}$ slab as $\sigma_{I_j}^{2}$ and $r_{0_j}$, respectively. To accurately model the atmospheric channel using multiple slabs with a phase screen located in the midway of each slab, we meet the two conditions described in~\cite{Martin88} (i.e. $\sigma_{I_j}^{2}\!<\!0.1$ and $\sigma_{I_j}^{2}\!<\!0.1\sigma_{I}^2$) by setting $N_S$ and $h_j$ through a numerical search.
A schematic illustration of our channel modeling (with $N_S=6$) is provided in Fig.~\ref{fig:modelS}(b). In our simulations $N_S$ ranges from $6$ to $12$ depending on specific settings.

After determining the widths of the atmospheric slabs, the realizations of the corresponding phase screens are generated using the Fast-Fourier-Transform (FFT)-based spectral domain algorithm~\cite{McGlamery67}. This method involves the filtering of a complex Gaussian random field using the phase power spectral density (PSD) function of the atmospheric turbulence. In this paper, we adopt the modified von Karman model, giving the phase PSD function for the $j^{\text{th}}$ slab
\begin{equation}\label{mvKPSD}
\Phi_{\phi_j}^{\text{mvK}}(f)=0.023 r_{0_j}^{-5 / 3} \frac{\exp \left(-f^{2} / f_{m}^{2}\right)}{\left(f^{2}+f_{0}^{2}\right)^{11 / 6}},
\end{equation}
where $f$ is the magnitude of the two-dimensional spatial frequency vector in the transverse plane in cycles/m, $f_0 = 1/L_{\text{outer}}$, and $f_m= 0.9422/l_{\text{inner}}$~\cite{SimAP}.


For the free-space propagation, we utilize the FFT-based angular spectrum method (for details of this method one can refer to e.g.~\cite{FourierOptics,SimAP}). In this study we utilize a physical optics propagation library named \textit{PROPER}~\cite{PROPER} to perform this method.


\subsubsection{Quantum state evolution}
To illustrate the undesirable decoherence effects caused by the atmospheric turbulence, we formally describe the evolution of an OAM eigenstate within a satellite-to-Earth channel. Assuming that Alice sends a single-photon OAM eigenstate $|l_t\rangle$ to Bob's ground station through an atmospheric channel. Under one realization of the atmospheric channel, the evolution of such a single-photon OAM eigenstate can be described by a  unitary operator $U_{\text{turb}}(L)$~\cite{EntanglementProtectionAO2019}. Denoting the received state as $\left|\psi_{l_t}\right\rangle$, we have
\begin{equation}\label{Eq.OAM_SingleState}
\left|\psi_{l_t}\right\rangle=U_{\text{turb}}(L)\left|l_t\right\rangle.
\end{equation}
The received single-photon state can be expanded in the OAM basis as
\begin{equation}\label{Eq:RxStateSingle}
\left|\psi_{l_{t}}\right\rangle=\sum_{l} c_{l,l_{t}}(L)| l\rangle,
\end{equation}
where $c_{l,l_t}(L) =\langle l|U_{\text {turb}}(L)|l_t\rangle$.

\begin{figure}[hbt!]
	\centering
	\includegraphics[scale=0.22]{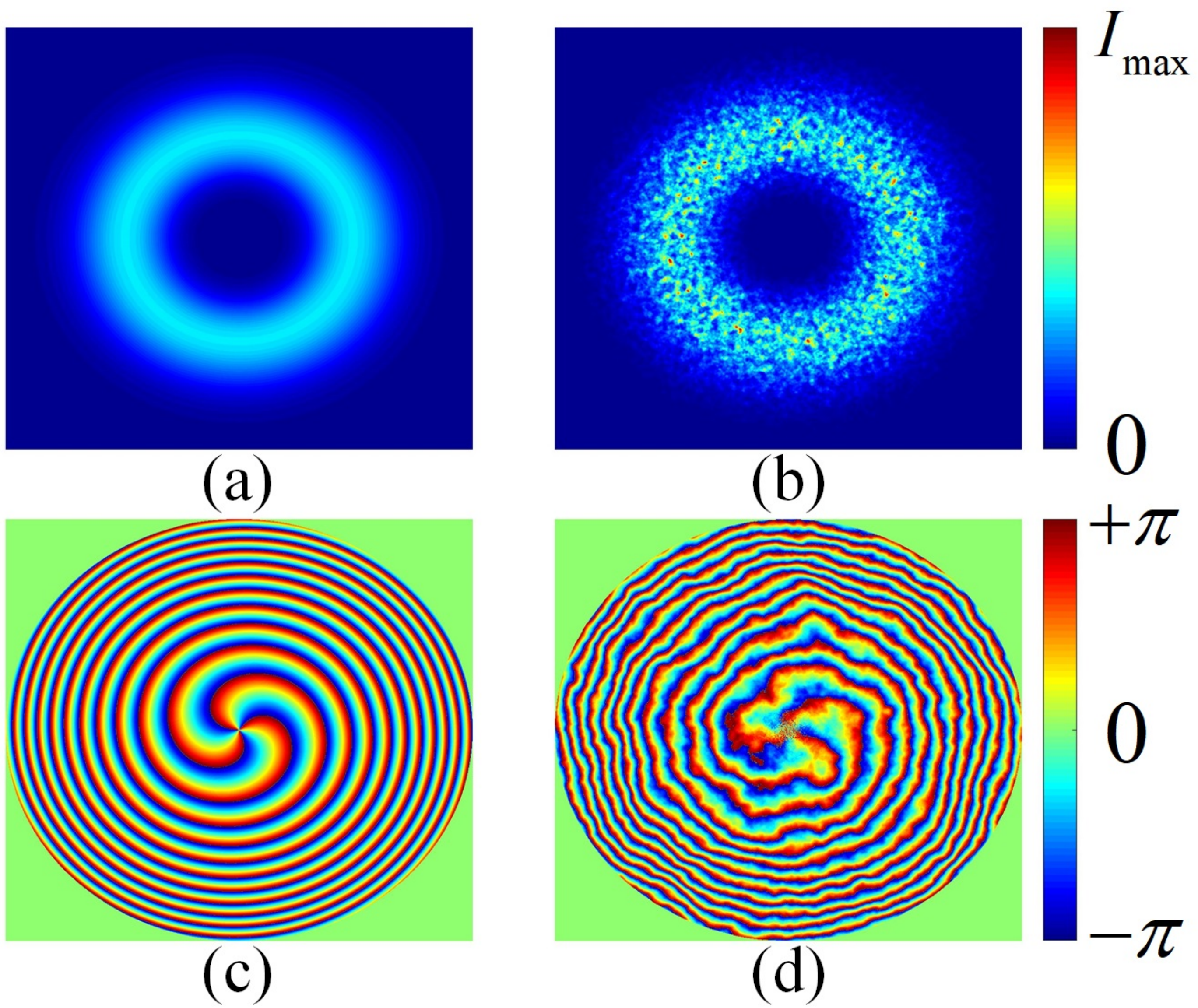}
	\caption{Intensity (a)(b) and phase (c)(d) profiles for an $\text{LG}_{03}$ beam after a vacuum propagation (left) and a realization of atmospheric propagation (right).}
	\label{fig:oamstates}
\end{figure}

In this work, the evolution of a single-photon OAM eigenstate is simulated by the atmospheric propagation of the corresponding classical $\text{LG}$ beam via the split-step method.
In Fig.~\ref{fig:oamstates} we plot the intensity and phase profiles of an $\text{LG}_{03}$ beam after vacuum propagation (i.e. propagation without atmospheric turbulence) and one realization of atmospheric propagation. After a vacuum propagation, we have $c_{l,l_t}\!=\!\delta_{l,l_t}$ due to the orthogonality of OAM eigenstates. After atmospheric propagation, however, the turbulence-induced distortions lead to crosstalk. At the receiver, $\left|\psi_{l_{t}}\right\rangle$ is generally a superposition of OAM eigenstates, and thus it is no longer orthogonal to any OAM eigenstate. The resulting crosstalk causes entanglement decay and thus degrades the performance of OAM-QKD.

\subsection{OAM-QKD over satellite-to-Earth channel}
\label{SubSec:OAM_QKD_Analyses}
Now let us analyze the performance of OAM-QKD protocols introduced in Section~\ref{SubSec:Gen_QKD_Protocol} over the satellite-to-Earth channel.
Specifically, we are interested in the achievable QKD performance over the satellite-to-Earth channel. The QKD performance is quantified by the secret key rate $K$ in \textit{bits per sent photon}, and throughout this work we use the unit \textit{bits per photon} for short.
 
For a $d$-dimensional OAM-QKD protocol, Alice generates OAM-photon pairs, each pair being in the maximally entangled state
\begin{equation}\label{Eq:IniEntStateG}
\ket{\Phi_0}=\sum_{l_t \in \mathcal{H}_d} \frac{1}{\sqrt{d}}\ket{l_t}\ket{l_t},
\end{equation}
where $\mathcal{H}_d$ is a $d$-dimensional encoding subspace. From each pair one photon is sent to Bob through a satellite-to-Earth quantum channel. At the output, the quantum state shared between Alice and Bob \textit{before any measurement} is given by
\begin{equation}\label{Eq.RxBiPhotonG}
\ket{\Phi_{\text{turb}}}=\{\mathbb{1} \otimes U_{\text{turb}}(L)\}\ket{\Phi_0}=\sum_{l_t \in \mathcal{H}_d}\sum_{l \in \mathcal{H}_{\infty}} \frac{c_{l,l_{t}}}{\sqrt{d}}\ket{l_t}\ket{l},
\end{equation}
where $\mathbb{1}$ denotes an identity operator acting on Alice's photon, and $\mathcal{H}_{\infty}$ denotes an infinite-dimensional Hilbert space.

Although the initial state $\ket{\Phi_0}$ can be considered as a finite-dimensional state living in the $\mathcal{H}_d \otimes \mathcal{H}_d$ subspace, due to the crosstalk it spreads over the entire infinite-dimensional Hilbert space.
Since a practical system can only utilize a finite-dimensional  encoding subspace, a necessary procedure is to project the output state $\ket{\Phi_{\text{turb}}}$ onto the original $\mathcal{H}_d \otimes \mathcal{H}_d$ subspace. This procedure is realized by a post-selection at Bob's side, giving a post-selected (and un-normalized) state
\begin{equation}\label{Eq:PS}
\ket{\Phi_{\text{ps}}}=(\mathbb{1} \otimes O)\ket{\Phi_{\text{turb}}},
\end{equation}
where $O$ is the filtering operator acting on Bob's photon. Since Bob has no information on $U_{\text{turb}}(L)$, his filtering operator $O$ is equal to $\Pi_d=\sum_{l_p\in \mathcal{H}_d} {\ket{l_p}\bra{l_p}}$. By setting $O=\Pi_d$ the post-selected state in Eq.~(\ref{Eq:PS}) can be explicitly given as\footnote{We will later show in Section~\ref{SubSec:D} that  with channel information available Bob can adopt a different choice of $O$, giving a different form of $\ket{\Phi_{\text{ps}}}$.}
\begin{equation}\label{Eq:PS_nonD_General}
\ket{\Phi_{\text{ps}}}=(\mathbb{1} \otimes \Pi_d)\ket{\Phi_{\text{turb}}}=\sum_{l_t \in \mathcal{H}_d}\sum_{l \in \mathcal{H}_d} \frac{c_{l,l_{t}}}{\sqrt{d}}\ket{l_t}\ket{l}.
\end{equation}
Note that $\ket{\Phi_{\text{ps}}}$ is not normalized. In fact, the atmospheric propagation and the post-selection together form a completely positive (and non-trace-preserving) map $\Pi_{d}U_{\text{turb}}(L)$. It is obvious that the post-selection results in a loss of photons. However, this operation will not give Eve any information, since the lost photons are simply discarded by Alice and Bob and will not be used in key generation.

Since we are not interested in any specific realization of the atmospheric channel, we perform an ensemble average of $\ket{\Phi_{\text{ps}}}$ over different channel realizations. After averaging over channel realizations and performing re-normalization, the averaged state shared between Alice and Bob can be given as a mixed state described by
\begin{equation}\label{Eq.avg_state}
\rho_{AB}=\frac{\ensavg {\ket{\Phi_{\text{ps}}} \bra{\Phi_{\text{ps}}}}}{\mathcal{T}},
\end{equation}
where $\ensavg{\cdots}$ denotes an ensemble average, and $\mathcal{T}=\operatorname{tr}\left(\ensavg {\ket{\Phi_{\text{ps}}} \bra{\Phi_{\text{ps}}}}\right)$ is the trace required for re-normalization. Note that $\mathcal{T}$ quantifies the \textit{photon survival fraction} after post-selection.

Now we briefly recall how security analysis is performed and how key rate is calculated for our OAM-QKD protocols (for a complete and rigorous security analysis, one can refer to~\cite{Qudit_SecurityProof2010,Qudit_SecurityProof2012}). Utilizing the photons that survive the post-selection, the secret key rate $K_{1}$ can be expressed as\footnote{Note, here the key rate is per photon actually used in the key generation. For every photon sent (transmitted) it can be 'lost' either by being not hitting the receiver, or by not being post-selected via the projection operation. The parameter $\mathcal{T}$ states the fraction of sent photons that survive both these loss events. Effectively, the finite-sized receiver aperture is absorbed into the post-selection process.}
\begin{equation}\label{Eq:KeyRate}
K_{1}=I(A\!:\!B)-\chi(A\!:\!E),
\end{equation}
where $I(A\!:\!B)$ is the classical mutual information between Alice and Bob, and $\chi(A\!:\!E)$ is quantum information between Alice and Eve. Considering the fact that Eve holds a purification of $\rho_{AB}$, $\chi(A\!:\!E)$ can be explicitly given as
\begin{equation}\label{Eq:ChiXE}
\chi(A\!:\!E)=S(\rho_{AB})-\sum_{a} p(a) S(\rho_{B}|{a}),
\end{equation}
where $S(\cdot)$ denotes the von Neumann entropy, $a=0,\cdots,d-1$ denotes Alice's measurement outcome, $p(a)$ denotes the probability distribution of $a$, and $\rho_{B}|{a}$ is the state of Bob's photon conditioned on $a$. In the security analysis it is assumed that all errors are caused by Eve's eavesdropping attempts. The average error rate $Q$ can be expressed as
\begin{equation}\label{AVER}
Q=\frac{1}{N_B} \sum_{\beta=1}^{N_B}\sum_{\begin{subarray}{c} s,s' \\ s \ne s' \end{subarray}} \operatorname{tr}\left(\ket{\xi_{(\beta,s)}^{*}}\bra{\xi_{(\beta,s)}^{*}}\otimes\ket{\xi_{(\beta,s')}}\bra{\xi_{(\beta,s')}} \rho_{AB}\right).
\end{equation}
	
Starting from Eq.~(\ref{Eq:KeyRate}), (\ref{Eq:ChiXE}), $K_{1}$ is found to be a function of $Q$~\cite{Qudit_SecurityProof2010,Qudit_SecurityProof2012}.
For a $d$-dimensional QKD protocol utilizing all $(d+1)$ MUBs, $K_1$ can be calculated as
\begin{equation}\label{Eq:KeyRate_d+1MUB}
\begin{aligned} K_1=& \log _{2} d+\frac{d+1}{d} Q \log _{2}\left(\frac{Q}{d(d-1)}\right) \\ &+\left(1-\frac{d+1}{d} Q\right) \log _{2}\left(1-\frac{d+1}{d} Q\right). \end{aligned}
\end{equation}
Recalling a non-unity photon survival fraction $\mathcal{T}$, the achievable secret key rate $K$ is given by
\begin{equation}\label{Eq:FinalKey}
K = \mathcal{T}\times K_1.
\end{equation}

\section{Numerical Evaluation of QKD Performance}\label{Sec:Simulation}
In this section, we numerically evaluate the performance of the satellite-to-Earth OAM-QKD protocols analyzed in Section~\ref{SubSec:OAM_QKD_Analyses}. We carry out Monte Carlo simulations to numerically evaluate the secret key rate $K$. First, we generate 4000 independent realizations of the satellite-to-Earth channel. For each channel realization we perform a series of atmospheric propagations using the split-step method to obtain a realization of $|\Psi_{\text{ps}}\rangle$ (see Eq.~(\ref{Eq:PS})). Afterwards, realizations of $|\Psi_{\text{ps}}\rangle$ are used to obtain $\mathcal{T}$ and $\rho_{AB}$ (see Eq.~(\ref{Eq.avg_state})). Then $Q$ is evaluated from $\rho_{AB}$ (see Eq.~(\ref{AVER})), and $K_1$ is then evaluated from $Q$ (see Eq.~(\ref{Eq:KeyRate_d+1MUB})). Finally, $K$ can be evaluated using $K_1$ and $\mathcal{T}$ (see Eq.~(\ref{Eq:FinalKey})).

\subsection{General settings}\label{SubSec:SimSettings}
We restrict ourselves to the case of a low-Earth-orbit (LEO) satellite with a maximum satellite altitude $H\!=\!500\,\text{km}$. We consider two zenith angles, $\theta_z\!=\!0\degree$ and $\theta_z\!=\!45\degree$, giving a maximum channel distance of $L\!\sim\!500\,\text{km}$ and $L\!\sim\!700\,\text{km}$, respectively. Note that a higher $H$ leads to worse performance. This is because the beam radius on atmospheric entry increases with $H$ (due to diffraction), and this in turn results in increased distortion on the beam. A larger $\theta_z$ also leads to  worse performance, since the photon travels a longer distance within the turbulent atmosphere. Unless otherwise stated, when we refer to our results we will mean for all considered $H$ values (i.e. from $200\,\text{km}$ to $500\,\text{km}$) and for all considered $\theta_z$ values (i.e. $0\degree$ and $45\degree$). Also, throughout this work QKD performances are compared at the same satellite altitudes under the same zenith angles.

For the atmospheric parameters, we set $A\!=\!9.6\!\times\! 10^{-14}\, \text{m}^{-2 / 3}$ which accords with a realistic setting adopted in~\cite{Submarine_CVQKD}. We set $V_g\!=\!3\,\text{m/s}$, giving a  value of $v_{\text{rms}}\!=\!21\,\text{m/s}$. We set $L_{\text{outer}}\!=\!5$ m and $l_{\text{inner}}\!=\!1$ cm for the atmospheric turbulence~\cite{book,LIDAR_SlantPath_Scales}. For the optical parameters, we set $\lambda\!=\!1064\,\text{nm}$ in accord with existing entanglement sources (e.g.~\cite{1064nmES}), and set $w_0$ to $15\,\text{cm}$.

The loss of signal at the receiver will be a function of several system parameters as well as the atmospheric conditions. The beam width at the receiver is critical in determining the loss of signal, and is largely dependent on system parameters such as transmitter aperture (sets the beam waist $w_0$), and the distance between the satellite and the ground station. For a given optical wavelength, a given beam width at the receiver, and a given channel distance, the transmitter beam waist required can be easily determined. However, in our calculations we simply set the beam waist (in effect the transmitter aperture).
To orientate ourselves, we note that the Micius satellite (which orbits at an altitude of about $500\,\text{km}$), with an aperture size of $0.3$~m, provided a beam width of $12\,\text{m}$ at ground level at a channel distance of $1200\,\text{km}$~\cite{SatQKD,ChinaSatEBQKD}. For a satellite altitude of $500\,\text{km}$, the smallest beam width at ground level we will have in our calculations will be $2.2\,\text{m}$, corresponding to a $1\,\text{dB}$ loss at a receiver aperture of $1$~m radius (for a sea-level receiver, and a zenith angle of $0\degree$, this corresponds to our $w_0=15\,\text{cm}$).

We perform all simulations using a numerical grid of $2048\!\times\!2048$ points with a spatial resolution\footnote{Note, the spatial resolution (i.e. the grid spacing in the transverse plane) could be adaptively varied along the propagation path to minimize numerical errors in FFT-based wave propagation methods~\cite{SimAP}. However, in this work we fix the spatial resolution throughout the simulation. To validate this we perform a vacuum propagation over the length of the channel and compare the resulting simulated beam profile with an independently-derived analytical profile at the same channel distance. Such a test is performed for all considered channel distances and for LG beams with all considered OAM numbers, and we find that all numerical errors are negligible. We also note, when no phase modulation is set at the phase screens, our simulation results give $U_{\text{turb}}(L)\!\!~=~\!\!\mathbb{1}$.} of $5\,\text{mm}$. In generating the random phase screens using the FFT-based method, 3 orders of subharmonics are added using the method introduced in~\cite{Lane_SH1992} to accurately represent the low-spatial-frequency components contributed by large-scale turbulent eddies.


\subsection{Ideal circumstances}\label{SubSec:SimSettings_I}
We first explore a rather ideal circumstance for the receiver.
Adopting all settings of Section~\ref{SubSec:SimSettings}, we initially set the ground-station altitude to $h_0\!=\!3000\,\text{m}$ to avoid the strong atmospheric turbulence near the sea level.
 We also first  adopt a large receiver aperture of $r_a=4\,\text{m}$,  thus providing a zero-loss scenario.

First we investigate the performances of 2-dimensional and 3-dimensional OAM-QKD for different $l_0$ values under such an ideal circumstance.
For 2-dimensional OAM-QKD, we find that a large $l_0$ value generally leads to a higher secret key rate. Under $\theta_z=0\degree$, positive key rates of 0.03, 0.05, 0.06 bits/photon can be achieved at $H=500\,\text{km}$ for $l_0=2,\,3,\,4$, respectively. Under $\theta_z=45\degree$, we observe a reduction in the secret key rate ranging from $70\%$ to $100\%$. Specifically, only $l_0=4$ leads to a positive key rate of $10^{-3}$ bits/photon at $H=500\,\text{km}$. For 3-dimensional OAM-QKD, we find that a larger $l_0$ value does not always lead to a higher secret key rate. Under $\theta_z=0\degree$, despite the observation that $l_0=1$ does not lead to any positive key rate, there is no significant correlation between the achievable secret key rate and $l_0$ for $l_0=2,3,4$. Moreover, no positive key rate can be achieved at $H=500\,\text{km}$ for any considered $l_0$ value. Under $\theta_z=45\degree$, we find that {no positive key rate} can be achieved by 3-dimensional OAM-QKD. By comparing the QKD performances, we find that the performance of 3-dimensional OAM-QKD is overall inferior to the performance of 2-dimensional OAM-QKD over the satellite-to-Earth channel for a given $l_0$ value.



We then compare the performances of OAM-QKD of dimensions ranging from 2 to 9 under $\theta_z=0\degree$, and find that the QKD performance decreases as the dimension increases\footnote{When OAM-QKD of a dimension larger than 3 is involved in a performance comparison, for each dimension $d$ we choose a specific encoding subspace $\mathcal{H}_d$ that maximizes the key rate.}. Specifically, we find that no performance advantage can be achieved, by OAM-QKD of any dimension, against 2-dimensional OAM-QKD at $H>200\,\text{km}$. Furthermore, we find that OAM-QKD of dimensions larger than 5 achieve no positive key rate at all considered satellite altitudes and under all considered zenith angles. No positive key rate can be achieved by OAM-QKD of dimensions larger than 2 under $\theta_z=45\degree$.

It is widely anticipated that the use of higher-dimensional QKD can improve noise resistance and lead to a higher secret key rate.
However, all the observations reported in this subsection clearly indicate that an increased dimension \textit{cannot} improve the performance of OAM-QKD over the satellite-to-Earth channel, even under the ideal circumstance.
This finding can be explained by the fact that the maximally OAM-entangled state of a higher dimension is less robust against turbulence (note that the similar phenomenon has been observed in e.g.~\cite{EntanglementProtectionAO2019}). This will lead to an increased error rate which can be large enough to nullify the advantage of higher-dimensional QKD. Therefore, a lower secret key rate is achieved in spite of a higher photon survival fraction (due to an enlarged encoding subspace).
In other words, the theoretical capacity advantage provided by increasing the dimension in OAM-QKD is negated by the atmospheric turbulence over the satellite-to-Earth channel.

\subsection{Realistic circumstances}\label{SubSec:SimSettings_R}
Now we extend our scope to more realistic circumstances. Specifically, we discuss the impact of loss, and a lower ground-station (receiver) altitude, on the feasibility of satellite-to-Earth OAM-QKD.
\subsubsection{Loss}\label{SubSec:Loss}
The main source of loss in a satellite-to-Earth channel is diffraction loss. Diffraction loss is \textit{state-dependent} in OAM-QKD since OAM eigenstates with different OAM numbers experience different amounts of diffraction~\cite{OAM_SDD_0}. In order to investigate the impact of loss on the feasibility of satellite-to-Earth OAM-QKD, we adopt all settings of Section~\ref{SubSec:SimSettings_I} except setting the radius of the receiver aperture to $r_a=1\,\text{m}$. At $H=500\,\text{km}$ and under $\theta_z=0\degree$, setting $r_a=1\,\text{m}$ gives losses of 1 dB, 3.4 dB, 6.9 dB, 11.3 dB, 16.7dB to OAM eigenstates with OAM numbers $0$, $1$, $2$, $3$, $4$, respectively. We then re-evaluate the performances of 2-dimensional and higher-dimensional OAM-QKD.

\begin{figure}[!hbtp]
	\centering
	\includegraphics[scale=0.45]{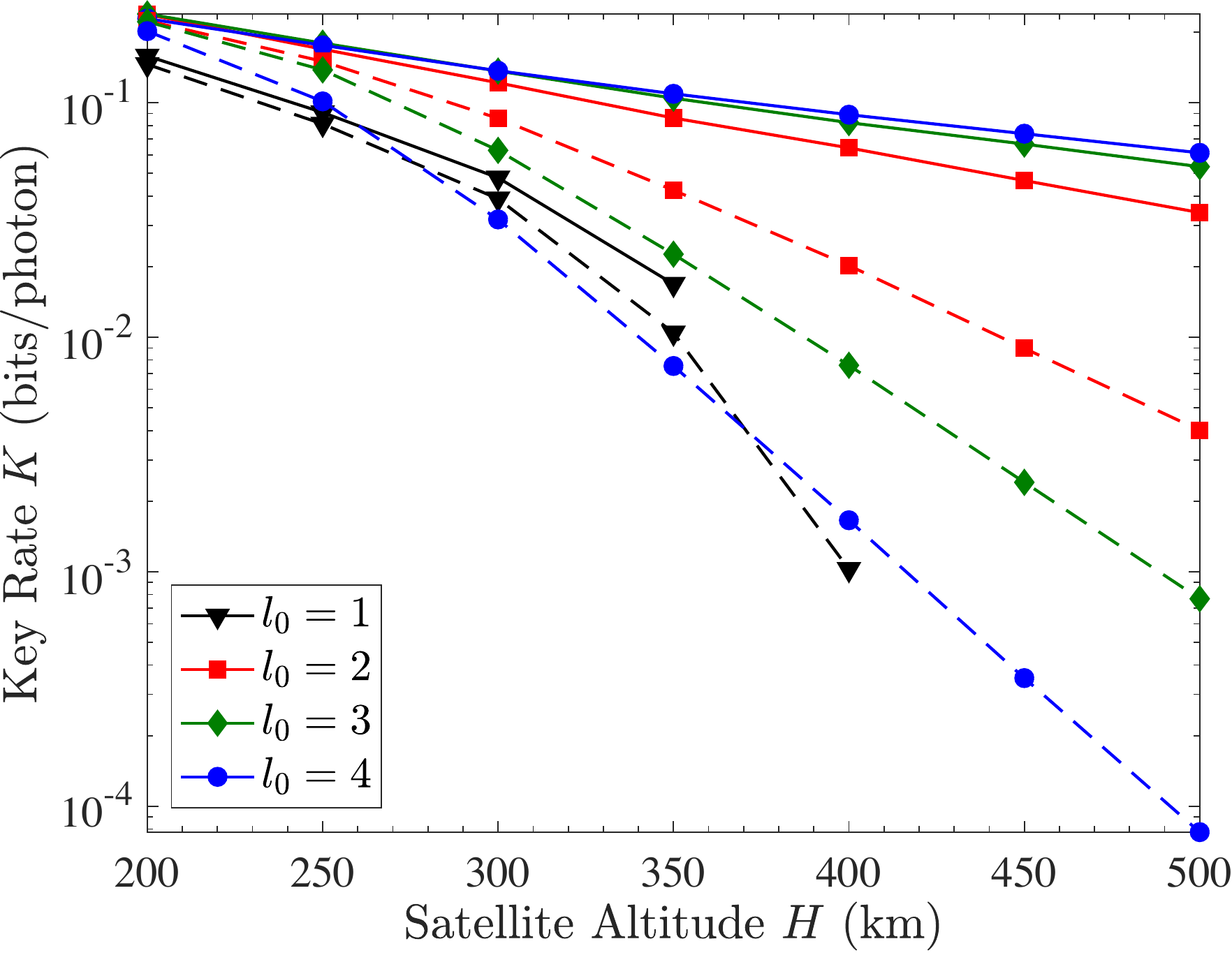}
	\caption{Secret key rates $K$ of 2-dimensional satellite-to-Earth OAM-QKD, achieved with $r_a=1\,\text{m}$ (dashed) and $r_a=4\,\text{m}$ (solid). These results are achieved under $h_0=3000\,\text{m}$ and $\theta_z=0\degree$.  Note that some curves end before reaching $H=500\,\text{km}$ due to a zero key rate. This happens when the average error rate $Q$ surpasses the tolerable error rate.}
	\label{fig:Loss}
\end{figure}

In Fig.~\ref{fig:Loss} we compare the performances of 2-dimensional OAM-QKD, achieved with $r_a=1\,\text{m}$ and $r_a=4\,\text{m}$ (zero loss), under $\theta_z=0\degree$ and $h_0=3000\,\text{m}$. From this figure, we  see that the loss degrades the performance of 2-dimensional OAM-QKD, and such a performance degradation is more significant for a larger $l_0$ value.

Higher-dimensional OAM-QKD is more sensitive to loss. For 3-dimensional OAM-QKD, after setting $r_a=1\,\text{m}$ we find that no positive key rate can be achieved at $H>300\,\text{km}$ under $\theta_z=0\degree$.
For OAM-QKD of dimensions larger than 3, setting $r_a=1\,\text{m}$ we find that no positive key rate can be achieved at $H>250\,\text{km}$.
Comparing the performances of OAM-QKD of different dimensions under loss, we find that 2-dimensional OAM-QKD is more robust against loss compared to higher-dimensional OAM-QKD. Indeed, the loss has a greater impact on higher-dimensional OAM-QKD due to its state-dependent nature (see related discussions in e.g.~\cite{OAM_SDD}).

\subsubsection{Receiver altitude}\label{SubSec:GndAlt}
\begin{figure}[!hbtp]
	\centering
	\includegraphics[scale=0.45]{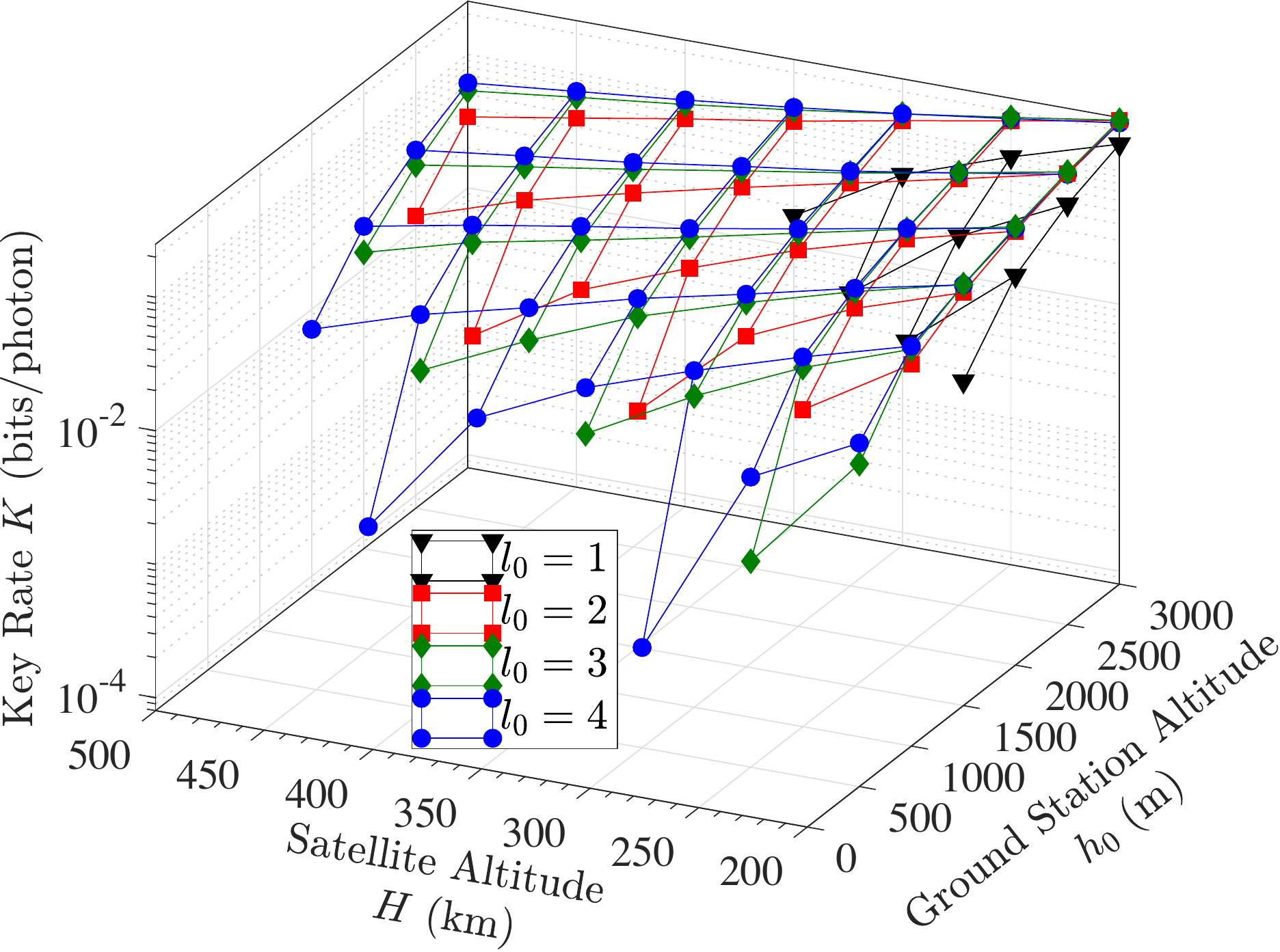}
	\caption{Secret key rates $K$ of 2-dimensional satellite-to-Earth OAM-QKD under different ground-station altitudes $h_0$. These results are achieved with $r_a=4\,\text{m}$ under $\theta_z=0\degree$. Again, some curves end before reaching $H=500\,\text{km}$ due to a zero key rate. }
	\label{fig:GndAlts}
\end{figure}

Lowering the ground-station altitude increases the turbulence strength, and intuitively this can degrade QKD performance. To see whether satellite-to-Earth OAM-QKD is feasible under lower ground-station altitudes, we adopt all settings of Section~\ref{SubSec:SimSettings_I} except for lower $h_0$ values.
In Fig.~\ref{fig:GndAlts} we compare the performances of 2-dimensional OAM-QKD, under $\theta_z=0\degree$, under different ground-station altitudes $h_0$. From this figure, we clearly see that the use of a lower ground-station altitude degrades the performance of 2-dimensional OAM-QKD at a given satellite altitude. We see that a positive key rate can still be achieved for $l_0=4$ at $H=500\,\text{km}$ under $h_0=1500\,\text{m}$. Under $\theta_z=45\degree$ we find that no positive key rate can be achieved at $H>300\,\text{km}$ under $h_0=2000\,\text{m}$.

OAM-QKD of a higher dimension is more sensitive to $h_0$. For 3-dimensional OAM-QKD, we find that a larger $l_0$ value is not more robust against performance degradation. We also find that, for $h_0=2000\,\text{m}$, no positive key rate can be achieved by 3-dimensional OAM-QKD at $H>250\,\text{km}$ even under $\theta_z=0\degree$.
For OAM-QKD of dimensions larger than 3, for $h_0=2000\,\text{m}$ we find that no positive key rate can be achieved at $H>200\,\text{km}$ even under $\theta_z=0\degree$.

\subsubsection{Sea-level receiver with reasonably-sized aperture}\label{SubSec:All}
Then we adopt all settings of Section~\ref{SubSec:SimSettings} and jointly set $r_a=1\,\text{m}$ and $h_0=0\,\text{m}$ to reflect a more realistic scenario where a sea-level receiver with a reasonably-sized aperture is used. Unfortunately, we find that no positive key rate can be achieved by OAM-QKD of any dimension, even under $\theta_z=0\degree$.

\section{Feasibility Through Channel Information}\label{Sec:Distillation}
In all previous sections we have assumed that channel knowledge is unavailable.
It is intuitive to think that channel information can be used to improve QKD performance. Indeed, in quantum communications a natural paradigm is to characterize the quantum channel through quantum process tomography (QPT)~\cite{SQPT,AAPT}, and cancel out the turbulence-induced effects accordingly~\cite{QCE_nonerp2017}. However, QPT is performed at the single-photon level, making the real-time characterization of quantum channels a challenging job. Recently it has been discovered that the state evolution of the classically entangled DoFs is equivalent to the state evolution of quantum entangled photons~\cite{QCE_nonerp2017}. Such an equivalence allows for the use of non-separable (i.e. DoF-entangled) states of classical light to characterize the quantum channel (e.g.~\cite{QCE_nonerp2017,OAM_SSP,HighD_QCEstimation_ClassicalLight2017,HighD_QCEstimation_ClassicalLight_CS2020}). By performing a state tomography on the output classical light, the quantum channel can be readily characterized in real time (for a comprehensive tutorial, see e.g.~\cite{VV_QST}).

In this section we explore the use of quantum channel information to improve the practical feasibility of satellite-to-Earth OAM-QKD. Inspired by~\cite{QCE_nonerp2017,OAM_SSP}, we utilize the quantum channel information acquired through a real-time quantum channel characterization utilizing non-separable states of classical light, and apply a quantum channel conjugation at the ground station. By quantum channel information we mean the Kraus operator of the channel, and by quantum channel conjugation we mean the application of a quantum conjugate filter that cancels out the turbulence-induced crosstalk.

\subsection{General method}\label{SubSec:D}
In this section, we demonstrate how a conjugate filter could be found, and we also analyze the impact of using that filter on OAM-QKD.
The non-separable states of classical light we use for channel characterization is given in the general form~\cite{HighD_QCEstimation_ClassicalLight2017}
\begin{equation}\label{Eq:IniCStateG}
\ket{\Phi_0^C}=\sum_{m}{\alpha_m}\ket{D^{(1)}_{m}}\ket{D^{(2)}_{m}},
\end{equation}
where the first DoF ${D}^{(1)}_m$ is a DoF that is not affected by the turbulent atmosphere (e.g. polarization or wavelength), the second DoF ${D}^{(2)}_m$ denotes the OAM DoF of light, $m$ indicates different basis elements in these DoFs, and $\alpha_m$ denotes the expansion coefficients such that $\sum_{m}{|\alpha_m|^2}=1$. We denote the encoding subspaces of ${D}^{(1)}_m$ and ${D}^{(2)}_m$ as $\mathcal{H}^{(1)}$ and $\mathcal{H}^{(2)}$, respectively. To faithfully characterize the quantum channel under study, it is required that $\dim(\mathcal{H}^{(1)}) = \dim({\mathcal{H}_d})$ and $\mathcal{H}^{(2)}=\mathcal{H}_d$.


In an OAM-QKD protocol Alice prepares OAM-photon pairs in the maximally entangled state $\ket{\Phi_0}$ described by Eq.~(\ref{Eq:IniEntStateG}). To help characterize the quantum channel Alice also generates classical light in the corresponding non-separable state $\ket{\Phi_0^C}$ described by Eq.~(\ref{Eq:IniCStateG}). While sending one photon of each entangled photon pair to Bob, Alice simultaneously sends the classical light through the same channel\footnote{We assume that the classical light is made orthogonal to the quantum signal using polarization or wavelength multiplexing techniques. For example, if the polarization (wavelength) DoF is used to construct the non-separable state in Eq.~(\ref{Eq:IniCStateG}), the wavelength (polarization) DoF should be used for multiplexing. We note that the turbulence effect on a propagating beam is wavelength-dependent, and this can potentially cause errors in quantum channel characterization. For simplicity, we assume that the wavelengths used for multiplexing and for constructing the non-separable state are chosen to be close enough to the wavelength of the quantum signal. Under such an assumption, the wavelength-dependent nature of the turbulence effect becomes negligible (see discussions in e.g.~\cite{HighD_QCEstimation_ClassicalLight2017}).}.
Since the state evolution of $\ket{\Phi_0^C}$ is equivalent to the state evolution of $\ket{\Phi_0}$, under a specific channel realization Bob can characterize the one-sided OAM quantum channel in the encoding subspace $\mathcal{H}_d$ by performing a state tomography on the received classical light. Specifically, Bob finds the Kraus operator $M$ that satisfies
\begin{equation}\label{Eq:SolvChannel}
(\mathbb{1} \otimes M)\ket{\Phi_0^C} = (\mathbb{1} \otimes\Pi_d)\ket{\Phi_{\text{turb}}^C},
\end{equation}
where $\ket{\Phi_{\text{turb}}^C}=(\mathbb{1} \otimes U_{\text{turb}}(L))\ket{\Phi_0^C}$ denotes the state of the received classical light at Bob's side. Note that the right-hand side of Eq.~(\ref{Eq:SolvChannel}) is known to Bob via his state tomography.

The Kraus operator $M$ can be expressed in its polar decomposition as
\begin{equation}\label{Eq:M}
M=U|M|,
\end{equation}
where $U$ is a unitary operator and $|M|=\sqrt{M^\dagger M}$ is a positive Hermitian operator. $|M|$ can be expressed in its spectral decomposition
\begin{equation}\label{Eq:|M|}
|M|=\sum_{j=1}^{d} {\gamma_{j}\ket{v_{j}}\bra{v_{j}}},
\end{equation}
where $\gamma_{j}$ and $\ket{v_{j}}$ denote the eigenvalues of $|M|$ and their corresponding eigenvectors, respectively. Note that $\ket{v_{j}}$ can be expressed as superpositions of the standard basis elements.

Considering the fact that $\gamma_{j}$ are smaller than 1, the conjugate filter cannot be directly constructed as $M^{-1}$. This is because $|M|^{-1}$ has eigenvalues larger than 1 and thus cannot be physically implemented due to a violation of the no-cloning theorem\footnote{One can show that, directly applying $M^{-1}$  leads to a \textit{noiseless amplification}. Such an operation is \textit{not} allowed in a deterministic fashion by the no-cloning theorem.}.
To construct a conjugate filter that does not violate the no-cloning theorem, inspired by the idea in~\cite{OAM_SSP} we consider a conjugate filter $\tilde{M}$ that achieves $\tilde{M}M\propto\mathbb{1}$. Specifically, we construct the conjugate filter as
\begin{equation}\label{Eq.Mt_General}
\tilde{M}=\left(\sum_{j=1}^{d}\frac{\gamma_{\text{min}}}{\gamma_{j}} \ket{v_{j}}\bra{v_{j}}\right) U^{\dagger}.
\end{equation}
where $\gamma_{\text{min}}=\min\{\gamma_{j}\, , j=1,\dots,d\}$.
Note that $\tilde{M}$ in Eq.~(\ref{Eq.Mt_General}) is a \textit{local Procrustean filter} which can be physically implemented (see experimental demonstrations of OAM Procrustean filters in e.g.~\cite{OAM_Concentration2003,OAM_entanglement12,OAM_Concentration2012}).

Under every channel realization, Bob constructs the Kraus operator $M$, constructs the conjugate filter $\tilde{M}$, and applies $\tilde{M}$ on his photon. Therefore, Alice and Bob share a post-selected state of the form
\begin{equation}\label{Eq:PS_conj_General}
\ket{\Phi_{\text{ps}}'}=(\mathbb{1} \otimes \tilde{M})\ket{\Phi_{\text{turb}}}=\gamma_{\text{min}}\ket{\Phi_{0}}.
\end{equation}
Note that $\ket{\Phi_{\text{ps}}'}$ is not normalized. From Eq.~(\ref{Eq:PS_conj_General}) it can be seen that the quantum channel conjugation results in a probabilistic entanglement distillation.  After averaging over channel realizations and performing re-normalization, $\rho_{AB}$ is now given by
\begin{equation}\label{Eq.avg_state_D}
\rho_{AB}'=\frac{\ensavg{\ket{\Phi_{\text{ps}}'}\bra{\Phi_{\text{ps}}'}}}{\mathcal{T}'}=\ket{\Phi_{0}}\bra{\Phi_{0}},
\end{equation}
where $\mathcal{T}'=\ensavg{\gamma_{\text{min}}^2}$ is the photon survival fraction when the quantum channel conjugation is applied. Note that $\mathcal{T}'$ can be interpreted as the probability of success of the quantum channel conjugation (which can be as low as $10^{-4}$ in extreme cases). Following the descriptions in Section~\ref{SubSec:OAM_QKD_Analyses}, $\rho_{AB}'$ and $\mathcal{T}'$ can be then used to evaluate the secret key rate $K$. It can be inferred from Eq.~(\ref{Eq.avg_state_D}) that $Q=0$ is achieved by the quantum channel conjugation at the cost of a low photon survival fraction.

Here we summarize the assumptions made, regarding the quantum channel characterization and the quantum channel conjugation, in this section. For simplicity, we assume that the channel Kraus operator $M$ is constructed without error (i.e. perfect state tomography on classical light), and the exact conjugate filter $\tilde{M}$ is applied to Bob's photon without error. Furthermore, these operations are assumed to be performed in real time. Such an assumption indicates that the time taken to perform a state tomography on the classical light is less than the coherence time of the atmospheric channel (typically on the order of milliseconds~\cite{book}).
Although no experiment has been demonstrated so far to indicate how fast such a state tomography can be done, in principle all the projective measurements required by such a state tomography can be done simultaneously with high signal-to-noise ratio. In addition, a recent work~\cite{SingleShot_VV} has demonstrated that a complete state tomography can be done in one shot if the classical light is in a pure state (note that this is valid under every specific channel realization).
We further notice that the paradigm adopted here resembles the concept of  Adaptive Optics (AO) where a servo loop system tracks (with a wavefront sensor) and corrects (with a deformable mirror) the turbulence effect in a real-time fashion. Given the fact that current AO systems have no significant trouble keeping up with the temporal evolution of turbulence, we believe that the quantum channel characterization and the following quantum channel conjugation can also be performed in real time.


\subsection{QKD performance with quantum channel conjugation}\label{Subsec:}
To numerically investigate the impact of the quantum channel conjugation on QKD performance, we  adopt the settings of Sections~\ref{SubSec:SimSettings_I},~\ref{SubSec:SimSettings_R} and re-evaluate the performances of satellite-to-Earth OAM-QKD of different dimensions.  In 2-dimensional OAM-QKD, we assume that the vector vortex beam is used for quantum channel characterization (for a detailed review on vector vortex beams one can refer to e.g.~\cite{VV_Rev}). Specifically, Eq.~(\ref{Eq:IniCStateG}) is explicitly given by
\begin{equation}\label{Eq:VectorVortex}
\ket{\Phi_0^C}=\frac{1}{\sqrt{2}}\left( \ket{R}\ket{l_0}+\ket{L}\ket{-l_0}\right) ,
\end{equation}
where $\ket{R},\ket{L}$ denote right and left circular polarization states, respectively.
In 3-dimensional OAM-QKD, the vector vortex beam cannot be used due to the constraint of the 2-dimensional Hilbert space imposed by the polarization DoF. It has been proposed in~\cite{HighD_QCEstimation_ClassicalLight2017} that the wavelength DoF of light is a promising candidate to replace the polarization DoF of light in $\ket{\Phi_0^C}$. Specifically, Eq.~(\ref{Eq:IniCStateG}) is explicitly given by
\begin{equation}\label{Eq:WL_OAM}
\ket{\Phi_0^C}=\frac{1}{\sqrt{3}}(\ket{\lambda_1}\ket{l_0}+\ket{\lambda_2}\ket{0}+\ket{\lambda_3}\ket{-l_0}),
\end{equation}
where $\lambda_m$ denote different wavelengths. Although no experiment has been demonstrated so far, the use of classical light in a state described by Eq.~(\ref{Eq:WL_OAM}) is theoretically feasible~\cite{HighD_QCEstimation_ClassicalLight2017}.
Since there is no fundamental limitation on dimension if the wavelength DoF is adopted, we also use this paradigm for quantum channel characterization in OAM-QKD of higher ($d>3$) dimensions.

First, we compare the QKD performances achieved with and without the quantum channel conjugation under the ideal circumstances in Section~\ref{SubSec:SimSettings_I}. We find that, with the help of the quantum channel conjugation, positive (and improved) secret key rates can be achieved by OAM-QKD of all considered dimensions at all considered satellite altitudes under all considered zenith angles. We also find that the use of smaller OAM numbers leads to a higher secret key rate. For 2-dimensional and 3-dimensional OAM-QKD, this means a smaller $l_0$ value leads to better performance. For OAM-QKD of dimensions larger than 3 this means using OAM numbers as small as possible to construct the encoding subspace leads to a better performance. Comparing the performances achieved by OAM-QKD of different dimensions, we find that an increased dimension can improve the performance of OAM-QKD over the satellite-to-Earth channel when the quantum channel conjugation is applied. Specifically, we find that 5-dimensional OAM-QKD  achieves the highest performance at $H\ge300\,\text{km}$ under $\theta_z=0\degree$. Under $\theta_z=45\degree$, 3-dimensional OAM-QKD achieves the highest performance at $H\ge250\,\text{km}$.

\begin{figure}[hbtp!]
	\centering
	\includegraphics[scale=0.45]{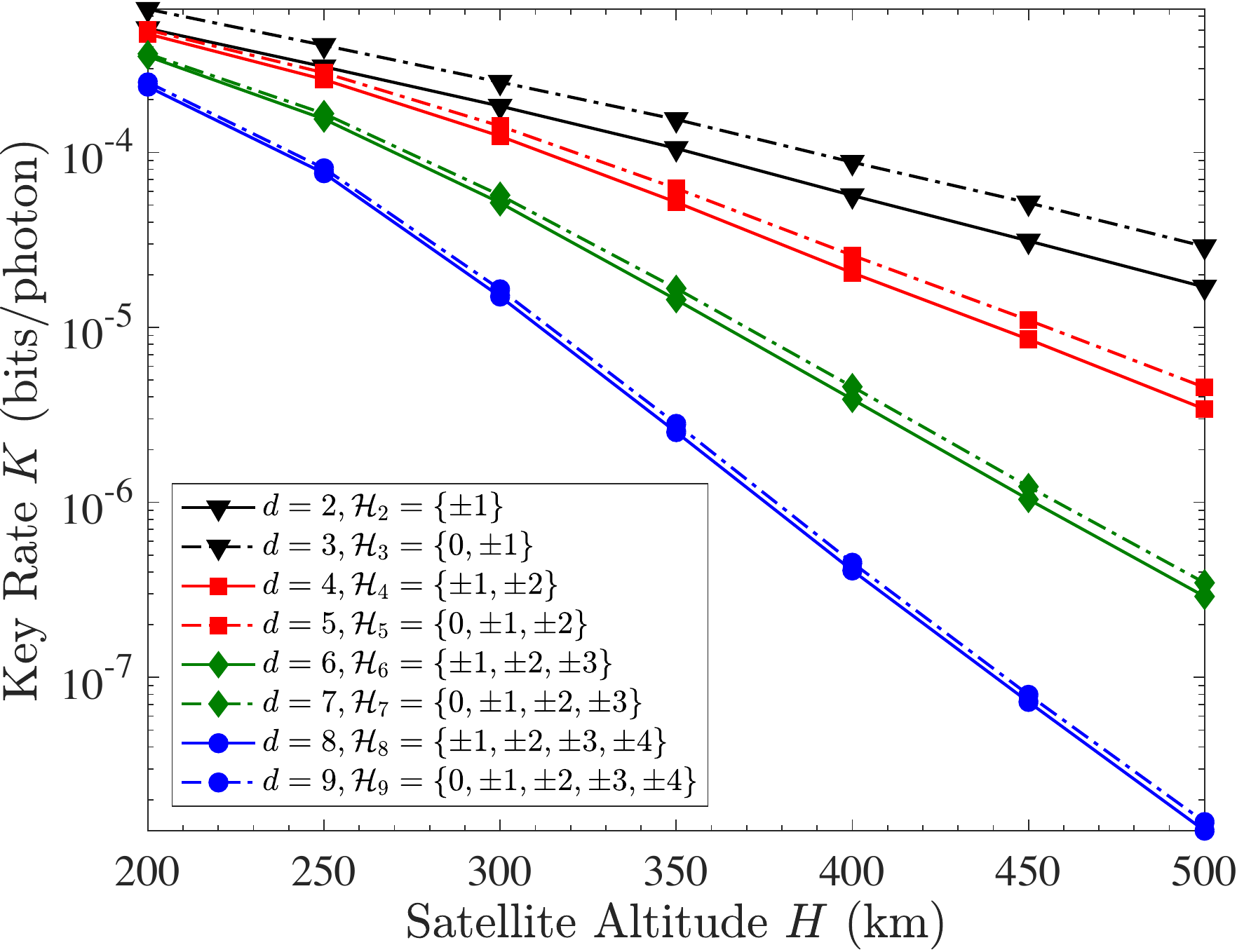}
	\caption{Performances of satellite-to-Earth OAM-QKD of different dimensions achieved with the quantum channel conjugation. These results are achieved with $r_a=1\,\text{m}$, under $h_0=0\,\text{m}$ and $\theta_z=45\degree$. A specific encoding subspace $\mathcal{H}_d$ is chosen to maximize the key rate for each dimension $d$.}
	\label{fig:ZA45_Wind21_h00_Rx100_Distillation_NP}
\end{figure}

Then, we adopt the settings in Section~\ref{SubSec:All} and evaluate the performance of OAM-QKD achieved with the quantum channel conjugation under the realistic circumstance where a sea-level (i.e. $h_0=0\,\text{m}$) ground station and a reasonably-sized (i.e. $r_a=1\,\text{m}$) receiver aperture is used. We find that, with the help of the quantum channel conjugation, positive secret key rates can be achieved by OAM-QKD of all considered dimensions. Such an observation not only holds under $\theta_z=0\degree$, but also holds under $\theta_z=45\degree$ (where a higher loss and more severe turbulence effect is anticipated). Specifically, in Fig.~\ref{fig:ZA45_Wind21_h00_Rx100_Distillation_NP} we present the QKD performances achieved with the quantum channel conjugation under $\theta_z=45\degree$. From this figure we can see that OAM-QKD of dimension 3 outperforms OAM-QKD of all other considered dimensions.

\begin{figure}[hbtp!]
	\centering
	\includegraphics[scale=0.45]{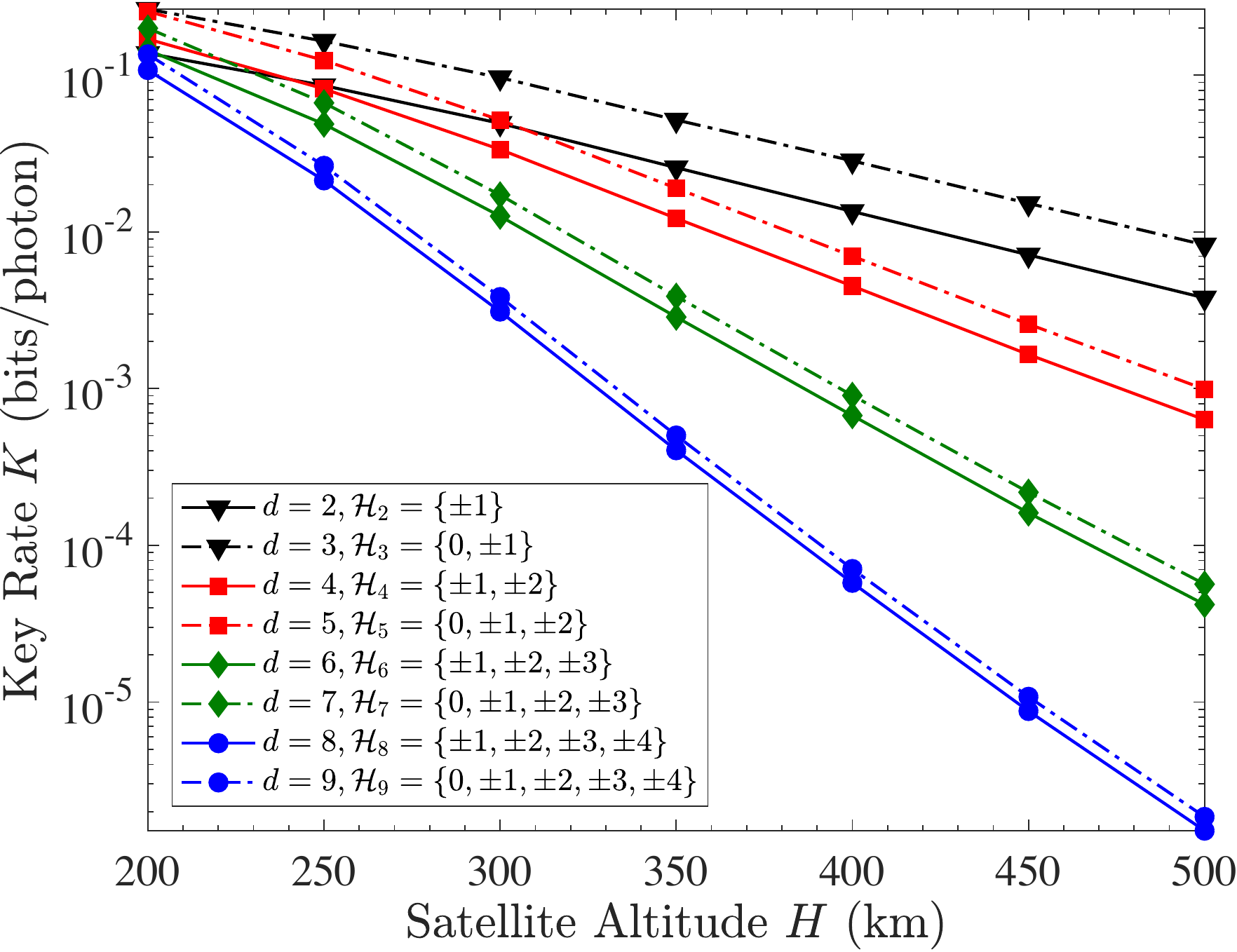}
	\caption{Performances of satellite-to-Earth OAM-QKD of different dimensions  achieved with the quantum channel conjugation. These results are achieved with $r_a=1\,\text{m}$, under $h_0=3000\,\text{m}$ and $\theta_z=45\degree$. A specific encoding subspace $\mathcal{H}_d$ is chosen to maximize the key rate for each dimension $d$.}
	\label{fig:Key_Wind21_Rx100_Distillation}
\end{figure}

Finally, we move to a better circumstance where a high-altitude ($h_0\!=\!3000\,\text{m}$) ground station with a reasonably-sized (i.e. $r_a\!=\!1\,\text{m}$) receiver aperture is used, and we evaluate the performance of OAM-QKD achieved with the quantum channel conjugation. In Fig.~\ref{fig:Key_Wind21_Rx100_Distillation} we plot the resulting QKD performances under $\theta_z\!=\!45\degree$. We first again find that the quantum channel conjugation leads to positive secret key rates for all dimensions at all considered satellite altitudes, and that 3-dimensional OAM-QKD achieves the highest QKD performance. Comparing the results in Fig.~\ref{fig:Key_Wind21_Rx100_Distillation} and Fig.~\ref{fig:ZA45_Wind21_h00_Rx100_Distillation_NP}, we clearly see the significant performance improvements provided by a higher ground-station altitude.

In summary, the quantum channel conjugation leads to positive (and improved) secret key rates at all considered satellite altitudes under all considered zenith angles, even under loss and a low ground-station altitude of $0\,\text{m}$. The quantum channel conjugation also enables the theoretically-predicted secret key rate advantage provided by an increased dimension in OAM-QKD over the satellite-to-Earth channel.


\subsection{Additional noise contributions}\label{Subsec2:}

We note that in our calculations we have assumed perfect  tomography implemented in real-time. Of course, in practice this perfect outcome can never be realized. The accuracy and  timescale for implementation of any tomography are a function of the specific measurements pursued and the number of signals analyzed e.g.~\cite{QT}. However, given the coherence timescale of the channel is of order a millisecond, and that we are using classical light as the probe (in effect no limit on sample size),
it could be anticipated that enough signals could be analyzed in real-time providing infidelities between the true and reconstructed quantum states less than $5\%$ \cite{QT}. The presence of tomography noise will manifest itself in our key rate calculations  through the design of a conjugate filter targeted at a different (erroneous) state, that ultimately leads to the state produced possessing less than maximal entanglement. The error rate $Q$, therefore, becomes non-zero which in turn impacts our final key rate (see Eq.~(\ref{Eq:KeyRate_d+1MUB})).

We also have assumed that our channel noise is entirely a consequence of phase perturbations and loss (the latter leading to vacuum contributions to the state). Although beam misalignment caused by turbulence-induced beam wander is negligible in the downlink from satellite to Earth, direction tracking errors in the transmitter and/or receiver may also cause misalignment (recall the satellite are in low orbit and moving across the sky in timescales of minutes).
The presence of beam misalignment will lead to additional cross-talk in the received state, which will manifests itself in our key rate equations through as smaller survival fraction in the measurement process (see Eq.~(\ref{Eq:FinalKey})).

\begin{figure}[hbtp!]
	\centering
	\includegraphics[scale=0.45]{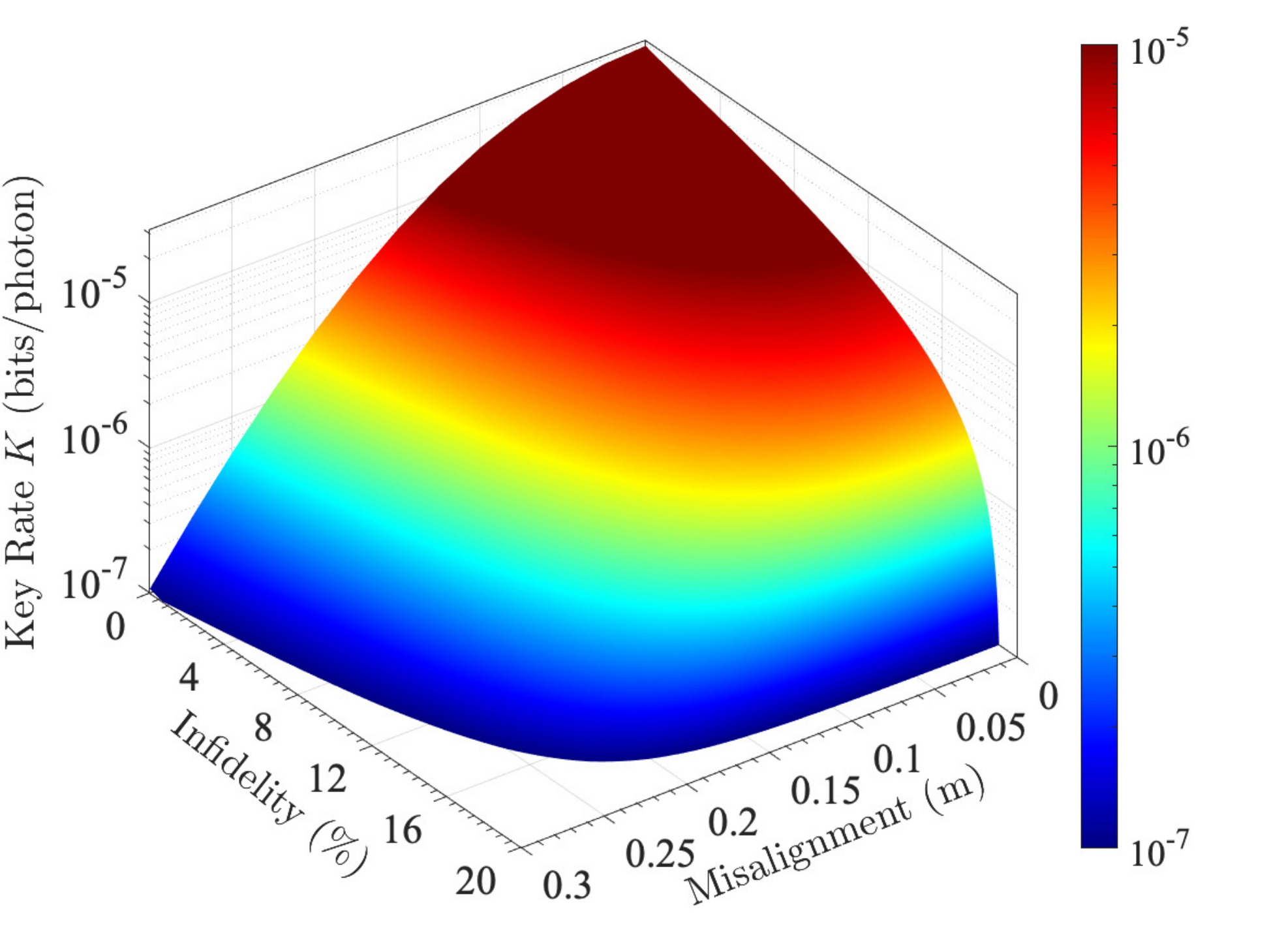}
	\caption{Performance of 3-dimensional satellite-to-Earth OAM-QKD achieved with the quantum channel conjugation, plotted against the channel conjugation error (in terms of infidelity) and misalignment. The satellite altitude is fixed to $H=500\,\text{km}$, and the results are achieved with $r_a=1\,\text{m}$, under $h_0=0\,\text{m}$ and $\theta_z=45\degree$. The encoding subspace $\mathcal{H}_3=\{-1,0,1\}$ is chosen to maximize the key rate.}
	\label{fig:Key_Wind21_Rx100_Error}
\end{figure}

In Fig.~\ref{fig:Key_Wind21_Rx100_Error} we illustrate the impact on our final key rate as a function of the additional noise terms discussed above.
Here, the performance of OAM-QKD is shown for a dimension of 3, and with the noisy channel conjugation and misalignment in the beam applied\footnote{The effect of misalignment is modeled as a deterministic misalignment operator acting on Bob's photon before the quantum channel conjugation. In other words, the magnitude (ranging from $0\,\text{m}$ to $0.3\,\text{m}$) and direction (fixed at $+45\degree$) of misalignment are constant under all channel realizations. The effect of a noisy channel conjugation is modeled as a deterministic depolarizing channel acting on Bob's photon after a perfect quantum channel conjugation, and the infidelity of such a depolarizing channel is used to quantify the channel conjugation error. We recognize our modeling of additional noise terms will not be an exact match to the real-world noise contributions.}. The settings are for a satellite altitude of $500\,\text{km}$, and a sea-level ground station with $r_a\!=\!1\,\text{m}$ under $\theta_z\!=\!45\degree$.
The considered 3-dimensional OAM-QKD protocol utilizes the encoding subspace $\mathcal{H}_3\!=\!\{-1,0,1\}$. The losses are $2.7\,\text{dB}$ and $7.4\,\text{dB}$ to OAM eigenstates with OAM numbers $0$ and $1$, respectively.
We can see that non-zero key rates can be found for a wide range of noise conditions. Beyond misalignment of $0.225\,\text{m}$ or infidelity of $18\%$ the key rate rapidly falls to less than $10^{-6}$ bits/photon. The region of nonzero key rates is at noise levels within current experimental reach. Note, that other detector noise components not explicitly mentioned, such as shot noise, dark counts and losses, are anticipated to be small relative to real-world misalignment noise, e.g.~\cite{PracticalCVRev}. Any additional detector noise can be readily mapped to an equivalent misalignment error of Fig.~\ref{fig:Key_Wind21_Rx100_Error}. Recall, that the classical signal is to be set by the system at much stronger intensity than the quantum signal. As such, most additional detector noise components can be made to have an impact on QKD rates well below the impact caused by a $0.05\,\text{m}$ misalignment error.

It should be noted that the quantum channel conjugation investigated in this work is not the only technique that can aid satellite-to-Earth OAM-QKD. It has been shown that AO techniques could improve the performance of OAM-based entanglement distribution in FSO channels (see e.g.~\cite{EntanglementProtectionAO2018,EntanglementProtectionAO2019}). AO techniques use a non-entangled classical light source as a probe, and their ability to negate turbulence heavily depends on the number of movable elements used for the required receiver-mirror deformation. But we note that phase perturbations across the transverse plane of the beam, when coupled to diffraction, leads to scintillation, and this cannot be completely negated by AO techniques. However, it is certainly the case that AO applied before any channel conjugation will only lead to improvement in the above results, particularly with regard to corrections of beam wander and direction tracking.
No report on the actual use of AO within the context of OAM entanglement distribution through long FSO channels is currently available.  In practice, we anticipate the channel conjugation method used here will lead to better negation of the atmospheric turbulence relative to AO, if either techniques is used on its own. However, further  research  on the coupling of quantum channel conjugation and advanced AO techniques may prove fruitful.

\section{Conclusions}\label{Sec:Conclusion}
The OAM of light has been considered as a promising DoF that gives access to a higher-dimensional Hilbert space, leading to potential higher capacity quantum communications.
In this work we explored the feasibility of performing satellite-to-Earth QKD utilizing the OAM of light. Specifically, we numerically investigated the performances of OAM-QKD of different dimensions achieved with different OAM numbers at different satellite altitudes $H$ under different zenith angles $\theta_z$. We found that utilizing the OAM of light in satellite-to-Earth QKD is indeed feasible between a LEO satellite and a high-altitude ground station.

First,  we considered an ideal circumstance where a high-altitude ground station  with a large receiver aperture (no loss) is used.
We then moved to less ideal circumstances and discussed the feasibility of satellite-to-Earth OAM-QKD under loss and a lower ground-station altitude.
However,
 we found that no positive secret key rate can be achieved at a sea-level ground station when a reasonable-sized aperture is used.
We then explored the use of quantum channel information as a means to improve the  feasibility of satellite-to-Earth OAM-QKD.
We assumed such information is acquired through a real-time quantum channel characterization utilizing non-separable states of classical light, and we used this information to perform a quantum channel conjugation  at the ground station.
We found that the quantum channel conjugation significantly improves the feasibility of OAM-QKD, and leads to positive secret key rates even under circumstances where a sea-level ground station with a reasonable-sized aperture is used.
We also found that the quantum channel conjugation enables a key rate advantage (provided by the higher dimensions of OAM-QKD) to be realized.


%

\end{document}